\documentclass[reprint,aip,superscriptaddress,showpacs,twocolumn]{revtex4-1}
\usepackage{graphicx}
\usepackage[ansinew]{inputenc}
\usepackage{amsfonts}
\usepackage[normalem]{ulem}
\usepackage{amsmath}
\usepackage{lipsum}
\usepackage{mathrsfs}
\usepackage{subfigure}
\usepackage{anysize}
\usepackage{color,graphicx}

\begin{document}
\title{Photovoltaic efficiency at maximum power of a quantum dot molecule}

\author{J. Lira}
\email{jefferson.santos@ufu.br}
\affiliation{Instituto de Física, Universidade Federal de Uberlândia, 38400-902, MG, Brazil}
\author{L. Sanz}
\email{lsanz@ufu.br}
\affiliation{Instituto de Física, Universidade Federal de Uberlândia, 38400-902, MG, Brazil}
\author{A. M. Alcalde}
\email{augusto.alcalde@ufu.br}
\affiliation{Instituto de Física, Universidade Federal de Uberlândia, 38400-902, MG, Brazil}

\date{\today}

\begin{abstract}
In this work, it is investigated the behavior of the efficiency at maximum power of a quantum dot molecule, acting as a device for photovoltaic conversion. A theoretical approach using a master equation, considering the effect of the energy offsets, and the width of the quantum barrier, identifies realistic physical conditions that enhance the photovoltaic response of the photocell. The results show the potentiality of increasing the gain in 30\% of maximum power delivered per molecule if compared with a single quantum dot. Also, the system exhibits gain when compared to the Chambadal-Novikov efficiency at maximum power, without exceeding Carnot's efficiency, as expected from the second law of thermodynamics.
\end{abstract}

\maketitle
A quantum heat engine (QHE) can be seen as a quantum version of the classical thermal machine, on a scale where nanoscopic effects can not be ruled out~\cite{benenti2017}. Nowadays, it has been suggested that energy conversion devices, such as photocells, can take advantage from non-classical properties, like quantum coherence between internal states or entanglement, to increase their photovoltaic response~\cite{scully2011}. These quantum effects are used as resources to maximize the power delivered by the device. The operating regime where maximum power is obtained is not necessarily the same where maximum efficiency is achieved. The efficiency at maximum power (EMP) is a more relevant measure than the maximum efficiency itself, since the latter can be obtained even when the power is zero~\cite{curzon1975,abah2014,esposito2010}. The EMP should be influenced by the various quantum resources directly related to the increase in power~\cite{svidzinsky2011}, in particular quantum coherence.

The state of the art of single and multiple quantum dot growth technology allows precise control of its structural characteristics, through which both, the electrical and optical responses, can be accurately controlled. In particular, tunneling coupled quantum dots also known as a quantum dot molecule (QDM) have been extensively studied and the controllability of their physical properties and interactions has been demonstrated~\cite{Stinaff06,Bracker2006,villas2004,krenner2005}. The flexibility in the control and design of their interactions turns them into promising systems for the implementation of quantum technologies in information processing and photovoltaic conversion~\cite{lent1997,li2001,scully2010,polman2016}. On the other hand, quantum coherence has been identified as an important performance enhancement resource for continuous heat engines~\cite{agarwalla2017,chen2016}. The latter effects were explored in the context of quantum coherence in the bath-system interactions. It has been shown that noise and an external source could induce quantum coherence, which can increase the power of a solar cell~\cite{scully2010,scully2011,dorfman2011} and is responsible for the highly efficient transfer of energy in photosynthetic systems~\cite{dorfman2013,dorfman2018}, which was confirmed in experimental studies of polymer solar cells~\cite{bittner2014}. One of the valuable resources for improving the efficiency of photovoltaic conversion in nanodevices is quantum coherence, which in turn is sensitive to the alignments of the conduction bands and valence.

The Carnot limit is reached by ideal thermal engines that operate reversibly between two thermal baths (cold and hot) so that the total entropy remains unchanged over a cycle. In 1961 Shockley and Queisser~\cite{shockley1961} showed that the maximum efficiency of a single $p-n$ junction solar cell would be approximately 44\% for a gap of 1.14 eV, if there is no loss of radioactive recombination. Other work~\cite{luque1997} reports that a cell with an intermediate band interval has an efficiency limit higher than that pointed out by Shockley, equivalent to 63\%. More recently, the Carnot efficiency has been challenged and several models of cyclic engines powered by quantum baths have been suggested~\cite{niedenzu2018,huang2012,abah2014,manzano2016,klaers2017}. In this scenario, it was demonstrated that, in fact, quantum engines that use squeezed thermal reservoirs, behave as more efficient thermal engines. However, its maximum efficiency can not be determined by the reversibility condition used by Carnot, as it results in an unattainable efficiency limit~\cite{niedenzu2018,abah2014,rossnagel2014,agarwalla2017,de2019}.

In this work, using a Markovian quantum master equation, it is explored the effect of the manipulation of band-offsets and tunneling coupling in a semiconductor QDM, to optimize the \emph{maximum delivered power} and the EMP provided by the quantum device. The last is studied as a quantifier of the photovoltaic characteristics of the system, once it allows observing the balance between the two fundamental quantities, efficiency, and power.

\begin{figure}
\centering
\includegraphics[scale=0.45]{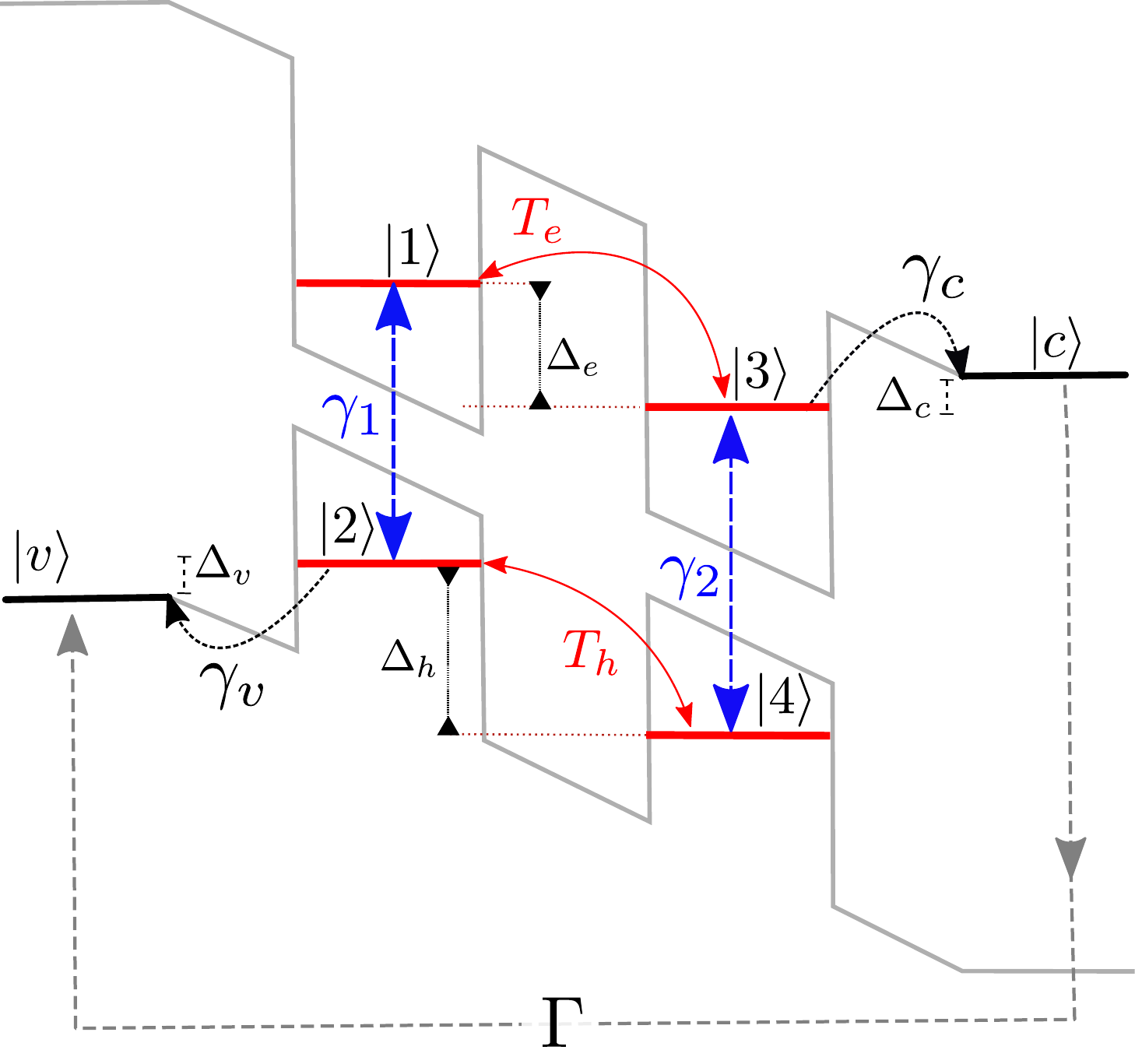}
\caption{A quantum dot molecule as a photocell: the red solid lines represent the energy levels of the QDM, while the black solid lines ($c$ and $ v$) are the conduction, $\left|c\right\rangle$, and valence, $\left|v\right\rangle$, states of the contacts. Tunneling couplings $T_{e(h)}$ are represented by red solid arrows. The energy offsets $\Delta_{e(h)}$ are shown using black dotted lines and $\Delta_{v(c)}$ by black dashed lines. The blue dashed and black dashed arrows represent incoherent couplings with solar radiation and ambient phonons respectively. Grey dashed arrow shows the process of relaxation with rate $\Gamma$ from $\left|c\right\rangle$ to $\left|v\right\rangle$. See text for more details.}
\label{fig1}
\end{figure}

A schematic representation of the system of interest is presented in Fig.~\ref{fig1}, being a semiconductor QDM of InAs/GaAs~\cite{Bracker2006} composed of two vertically aligned QDs separated by a tunneling barrier of width $d$ interacting with solar light, modeled as a bath of harmonic oscillators. We consider two optical transitions (blue dashed lines), between energy states labeled as $\left|1\right\rangle$, and $\left|2\right\rangle$ in the first QD, and $\left|3\right\rangle$, and $\left|4\right\rangle$, in the second QD. Coherent tunneling (red solid arrows) connects the states $\left|1\right\rangle$ and $\left|3\right\rangle$ ($\left|2\right\rangle$ and $\left|4\right\rangle$), with coupling strength $T_e$ ($T_h$) which depends on the barrier size $d$. The nanostructures are embedded in doped semiconductors represented by the contact levels $\left|c\right\rangle$ and $\left|v\right\rangle$. The energy offsets are modified in the calculations through a control parameter $\delta$ considering $\Delta_{e} = \Delta_{e}^0 + \delta$ and $\Delta_{h} = \Delta_{h}^0 - \delta$ where $\Delta_e^0 = \Delta_h^0 = 3$ meV, and $\Delta_c = \Delta_v = 2$ meV. The solar radiation pumps hot photons at a temperature $T_S$, an incoherent process (indicated by the blue arrows) that promotes electrons from the valence bands, $\left|2\right\rangle$ and $\left|4\right\rangle$, to the conduction bands, $\left|1\right\rangle$ and $\left|3\right\rangle$. Also, the parameters $\gamma_1$ and $\gamma_2$ describe the decay rates associated with the electron-hole recombination process linked to the transitions $\left|1\right\rangle \leftrightarrow \left|2\right\rangle$ and $\left|3\right\rangle \leftrightarrow \left|4\right\rangle$, respectively, and we set $\gamma_1=\gamma_2$. The states of the QDM are coupled to the reservoirs (the contacts) through phononic coupling (black dashed arrows) at temperature $T_a$, with rates $\gamma_{c}$ and $\gamma_{v}$ of the conduction band $\left|c\right\rangle$ and valence $\left|v\right\rangle$ reservoir, respectively. The levels $\left|c\right\rangle$ and $\left|v\right\rangle$ are connected to a load, which is modeled as a decay at rate $\Gamma$.

In the interaction picture, considering the rotating-wave approximation (RWA), the hamiltonian of the system is given by
\begin{eqnarray}
\hat{V}(t)=\hat{V}_T(t)+\hat{V}^S_{th}(t) + \hat{V}^a_{th}(t),
\label{eq:interactions}
\end{eqnarray}
where the first term describes the coherent tunneling of carriers between QD levels being
\begin{equation}
\hat{V}_T(t)=T_ee^{i\omega_{13}t}\left|1\right\rangle\left\langle 3\right| + T_he^{i\omega_{24}t}\left|2\right\rangle\left\langle 4\right| + \mathrm{H.c.}.
\end{equation}
Here $T_e$ ($T_h$) is the electron (hole) tunneling coupling between conduction (valence) levels. The connection between the tunneling couplings and the barrier width $d$ is modeled by fitting the experimental data~\cite{Bracker2006} by decay exponentials of the type $T_{e(h)}\approx e^{-d/d_{e(h)}}$, where $d_e = 7.14$ nm and $d_h = 3.37$ nm.
The second term in Eq.(\ref{eq:interactions}) takes into account the interaction between the QDM and the solar radiation, modeled here as a photon reservoir at sun temperature $T_S$:
\begin{eqnarray}\nonumber
\hat{V}^S_{th}(t)&=&\hbar\sum_{k}\left[g_{1k}e^{i(\omega_{12}-\nu_k)t}\left|1\right\rangle\left\langle 2\right|\hat{a}_k\right.\\
&&\left.+ g_{2k}e^{i(\omega_{34}-\nu_k)t}\left|3\right\rangle\left\langle 4\right|\hat{a}_k + \mathrm{H.c.}\right]\ ,
\end{eqnarray}
with $g_{1k}$ and $g_{2k}$ being the coupling strengths relatives to transitions $\left|1\right\rangle \leftrightarrow \left|2\right\rangle$ and $\left|3\right\rangle \leftrightarrow \left|4\right\rangle$, respectively. Also, $\hat{a}_k$ is the bosonic annihilation operator for the k-th mode of the radiation field with frequency $\nu_k$, and $\omega_{ij} = (E_i - E_j)/\hbar$ is the frequency of the transition $\left|i\right\rangle \leftrightarrow \left|j\right\rangle$. The third term in Eq.(\ref{eq:interactions}) describes the interaction with the phonon reservoirs
\begin{eqnarray}\nonumber
\hat{V}^a_{th}(t)&=&\hbar\sum_{l}^{}(g_{l}e^{i(\omega_{3c}-\nu_l)t}\left|3\right\rangle\left\langle c\right|\hat{b}_l + \mathrm{H.c.})\\ \nonumber
& & + \hbar\sum_{m}(g_{m}e^{i(\omega_{v2}-\nu_m)t}\left|v\right\rangle\left\langle 2\right|\hat{b}_m + \mathrm{H.c.})\ ,\\
\end{eqnarray}
where $g_{l}$ and $g_{m}$ are the couplings for the transitions $\left|3\right\rangle \leftrightarrow \left|c\right\rangle$ and $\left|2\right\rangle \leftrightarrow \left|v\right\rangle$, respectively. Here, $\hat{b}_i$ with $i = l,m$ are the boson operators for thermal phonons with frequency $\nu_i$. Because the coupling between the quantum molecule and the ambient phonons is weak, the RWA describes well the physics of the problem.

The equation of motion for the density matrix is given by
\begin{eqnarray}\nonumber
\dot{\hat{\rho}}(t)&=&-\frac{i}{\hbar}\mathrm{tr}_R[\hat{V}(t),\hat{\rho}(t_0)\otimes\hat{\rho}_R(t_0)]\\\nonumber
& &-\frac{1}{\hbar^2}\mathrm{tr}_R\int_{t_0}^{t}dt'[\hat{V}(t),[\hat{V}(t'),\hat{\rho}(t')\otimes\hat{\rho}_R(t_0)]]\ ,\\\label{eq:master}
\end{eqnarray}
where the trace has been taken over the radiation and phonon reservoirs variables. To examine the behavior of the photovoltaic properties of the QDM, within the framework of Weisskopf-Wigner approximation~\cite{scully2010,svidzinsky2011}, we solve numerically the master equation, Eq.~(\ref{eq:master}), in the stationary regime where $\dot{\rho}=0$. The physical quantities in the description of the problem depends on the value of the density matrix terms $\rho_{ij}$: the current, for example, is defined as $j=e\Gamma\rho_{cc}$, with $\rho_{cc}$ being the population of the state $\left|c\right\rangle$.

A preliminary analysis shows the manipulation of the energy offsets affects the power provided by the photocell considering different values of the barrier width $d$~\cite{Lira21a}. Here, because high power does not always correspond to high efficiency, the focus is to analyze the proper adjustment between photovoltaic conversion efficiency and delivered power. Once $\left|c\right\rangle$ and $ \left|v\right\rangle $ are connected by the load $\Gamma$, the voltage $V$, according to Fermi-Dirac statistics, produces $eV = E_c-E_v + k_BT_a \ln \left({\rho_{cc}}/{\rho_{vv}}\right)$, where $E_i$ is the state energy $i$ with an occupation $\rho_{ii}$ calculated in the steady state regime. The power delivered by the QDM photocell to the load is $P=jV$.
The \emph{maximum delivered power} $P_m$ is obtained by maximizing the power over the barrier width $d$ and the band offset control parameter. The ratio between $P_m$, and the incident solar radiation power~\cite{svidzinsky2011}, $P_S=jE_{12}/e$, defines the EMP as $\eta=P_m/P_S$.
\begin{figure}
	\centering
	\includegraphics[scale=0.6]{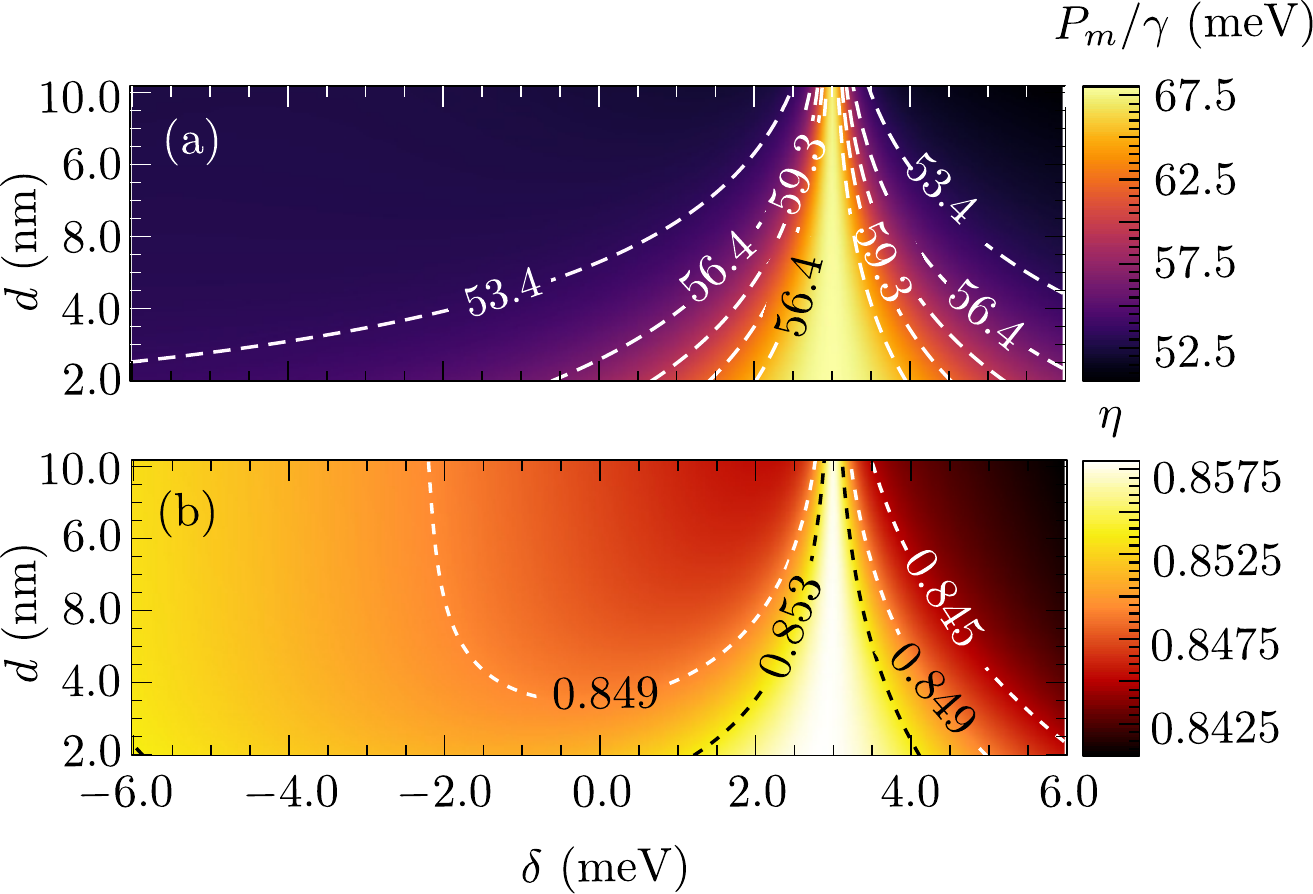}
	\caption{(a) Maximum delivered power $P_m$ and (b) the efficiency at maximum power (EMP) $\eta$ as functions of $d$, and the parameter $\delta$. The relaxation rates are considered in terms of the electron-hole recombination parameters as $\gamma_{c}=100\gamma$ and $\gamma_{v}=0.05\gamma$, where $\gamma=\gamma_1=\gamma_2$.}
	\label{Fig2}
\end{figure}
The Fig.~\ref{Fig2} shows the maximum power $P_m$ and the efficiency at maximum power $\eta$, as a function of the energy offset $\delta$ and barrier width $d$, for $\gamma_{v}=100\gamma$ and $\gamma_{c}= 0.05\gamma $~\cite{dorfman2011}. We chose realistic values of barrier separation given by $d =2-10$ nm, for which the tunneling coupling $ T_e $ varies between $ T_e=4.410-1.438$ meV ($T_h= 0.608-0.057$ meV). The behavior shows that the interplay between the coherent tunneling intensity and the tuning of the relevant energy levels establishes the efficiency gain attributed to the coherence. Furthermore,  the configuration $ \delta = 3 $ meV, where the valence levels of the QDM are in resonance ($ \Delta_e = 6 $ meV and $ \Delta_h = 0 $), provides the highest power and efficiency. Interestingly, this condition is practically independent of the barrier width $d$. However, the range of values $\delta$ with high power and efficiency $\delta$ values becomes broader as $d$ decreases and $T_{e,h}$ increases, which is interesting in terms of increasing the applicability of the QDM as a photovoltaic converter.

\begin{figure}
	\centering
	\includegraphics[scale=0.27]{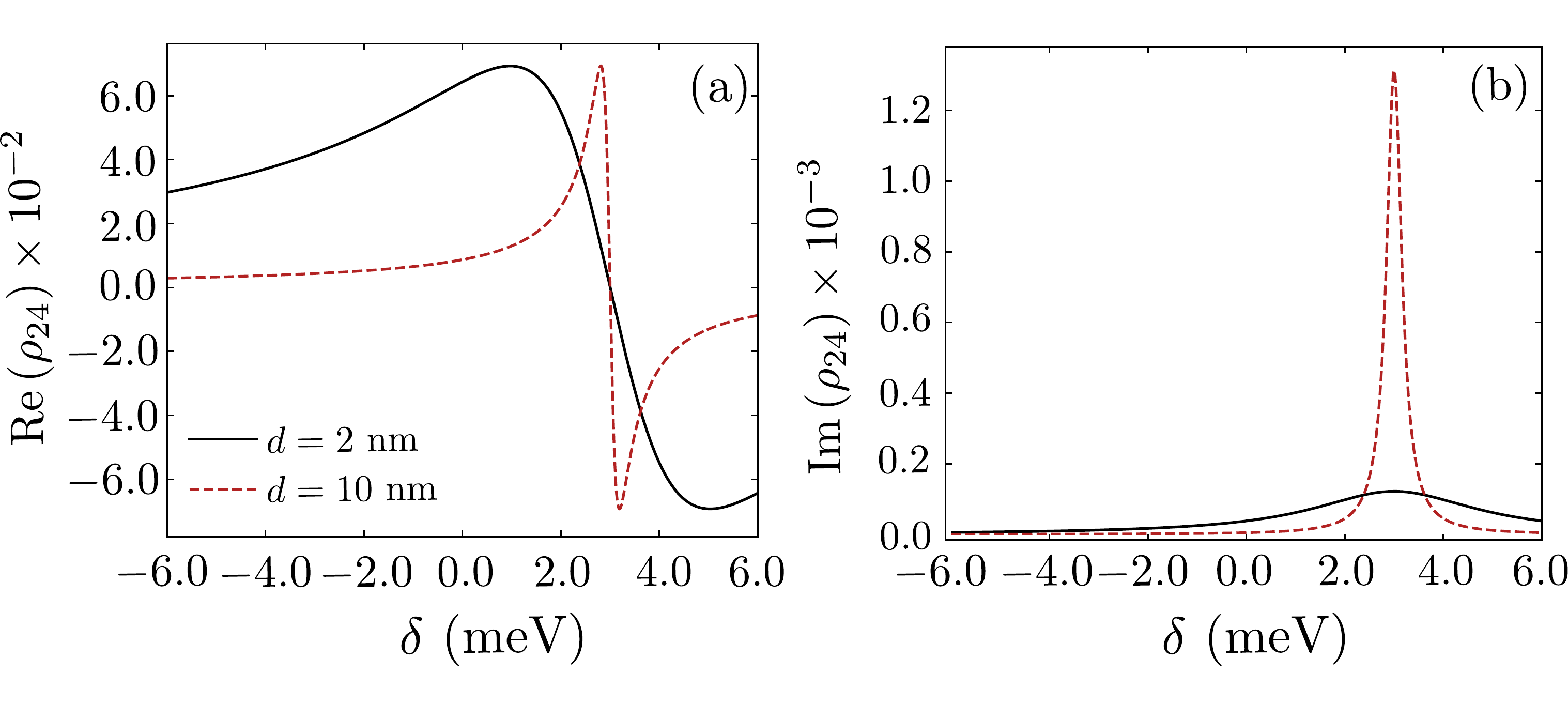}
	\caption{The coherence term $\rho_{24}$ in the maximum power point as a function of the energy offset parameter $\delta$. (a) Real part of the coherence term, $\mathrm{Re}\left(\rho_{24}\right)$, and (b) imaginary part $\mathrm{Im}\left(\rho_{24}\right)$, considering $d = 2$ (black solid line), and $d = 10$ nm (red dashed line).}
	\label{Fig3}
\end{figure}
The explanation behind the advantage of the condition $\delta=3$ meV emerges when it is analyzed the response of the system to the tunneling couplings $T_h$ and $T_e$. This is done by studying the behavior of the quantum coherences $\rho_{24}$ and $\rho_{13}$. The real and imaginary parts of both quantities were calculated as functions of $\delta$ and the barrier width $d$. Although $\rho_{13}$ (not shown here) does not have significant variations depending on both parameters, the behavior of the quantum coherence $\rho_{24}$ explains the photovoltaic behavior discussed before. Fig.~\ref{Fig3} illustrates the real part of $\rho_{24}$, panel (a), and the imaginary part, panel (b), as functions of the parameter $\delta$ for two different values of the barrier width $d$. The behavior of the imaginary part, panel (a), shows the regions where the maximum power and efficiency reach large values (see Fig.~\ref{Fig2}) are related with the range of values of $\delta$ where $\mathrm{Re}({\rho_{24}})$ varies from positive to negative values. Also, $\mathrm{Re}(\rho_{24})$ is null at $\delta=3$ meV, independently of the barrier width. From panel (b), the imaginary part shows a Lorentzian profile, whose width is proportional to $d$. Still, regardless of the value of $d$, the highest value is always obtained at $\delta=3$ meV. That means that at the resonance condition of the valence states, the population inversion occurs even for small coupling rates, with populations and coherences rapidly tending to the steady-state, making the photovoltaic conversion provided by QDM as efficient as possible.

\begin{figure}
	\centering
	\includegraphics[scale=0.42]{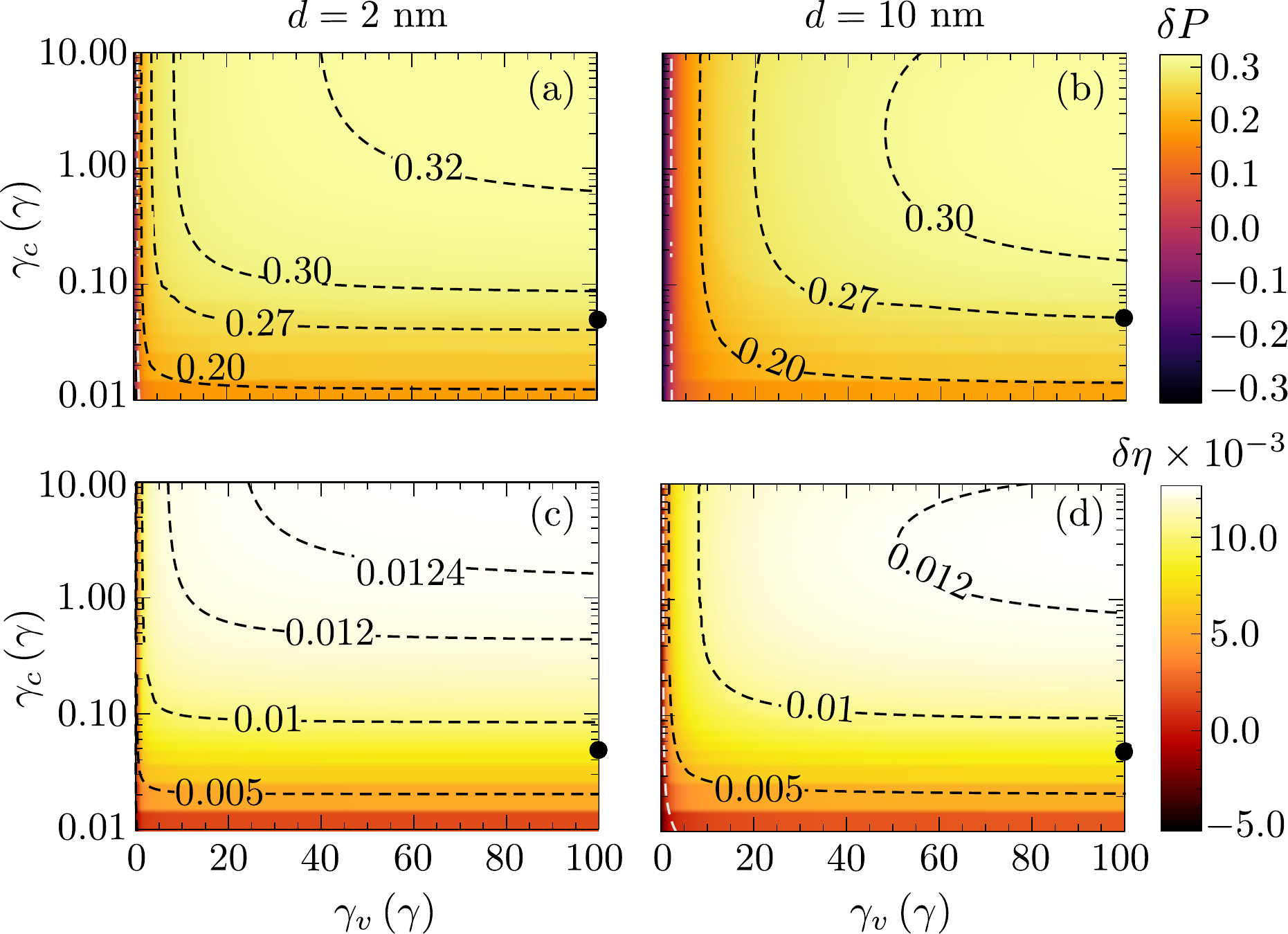}
	\caption{Relative variation of the power ($\delta P$), panels (a) and (b), and efficiency ($\delta \eta$), panels (c) and (d), as functions of the relaxation rates $\gamma_{c}$ and $\gamma_{v}$ considering $ \delta = 3 $ meV. For the left (right) panels, the barrier width is set in $ d = 2 $ nm ($10$ nm). The black dashed lines and text indicate some values for reference. Black points show the choice of relaxation parameters ($\gamma_{v}$, $\gamma_{c}$) considered in Fig.~\ref{Fig2}, and Fig.~\ref{Fig3}.}
\label{Fig4}
\end{figure}

At this point, it is interesting to quantify the advantage of using a QDM for photovoltaic conversion in relation to a single quantum dot (SDQ), as well as the role of relaxation rates. This is performed by using the relatives variation of the power and the efficiency, defined as $\delta X = (X^{\mathrm{QDM}}_{\mathrm{max}} - X^{\mathrm{SQD}})/{X^{\mathrm{SQD}}}$, where $X$ is either the power $P$ or the efficiency $\eta$. In Fig.~\ref{Fig4} the panels (a) and (b) show the behavior of the relative variation of the power, $\delta P$, while in (c) and (d) illustrate $\delta \eta$, both as functions of the relaxation rates $\gamma_{v}$ and $\gamma_{c}$. Two values of barrier width are considered: $d = 2$, panels on the left, and $d = 10$ nm, panels on the right. As a reference, the dashed lines highlight points in the plot where $\delta X > 0$ (QDM exceeds SQD). Dark colors remark conditions with $\delta X < 0$ (SQD exceeds QDM), and the black point on each plot indicates the choice of relaxation parameters $ \gamma_{v} = 100 \gamma $ and $ \gamma_{c} = 0.05 \gamma $~\cite{dorfman2011}, used in the previous figures. These results evidence the advantages of using QDM in actual applications of solar cells: for a large set of values of relaxation rates $\gamma_{v}$ and $\gamma_{c}$, the obtention of $\delta P > 0.3$ translates in a gain of 30\% in the maximum power delivered for the QDM, if compared to the SQD. The efficiency shows the same tendency, which becomes evident from the comparison of the bright regions of the panels corresponding to the same value of $d$. This gain is not heavily affected by changes of barrier width, if the energy offset corresponds to the optimal value of $\delta=3$ meV,  found from the analysis of the behavior of $P_m$ and $\eta$ discussed above

\begin{figure}
	\centering
	\includegraphics[scale=0.6]{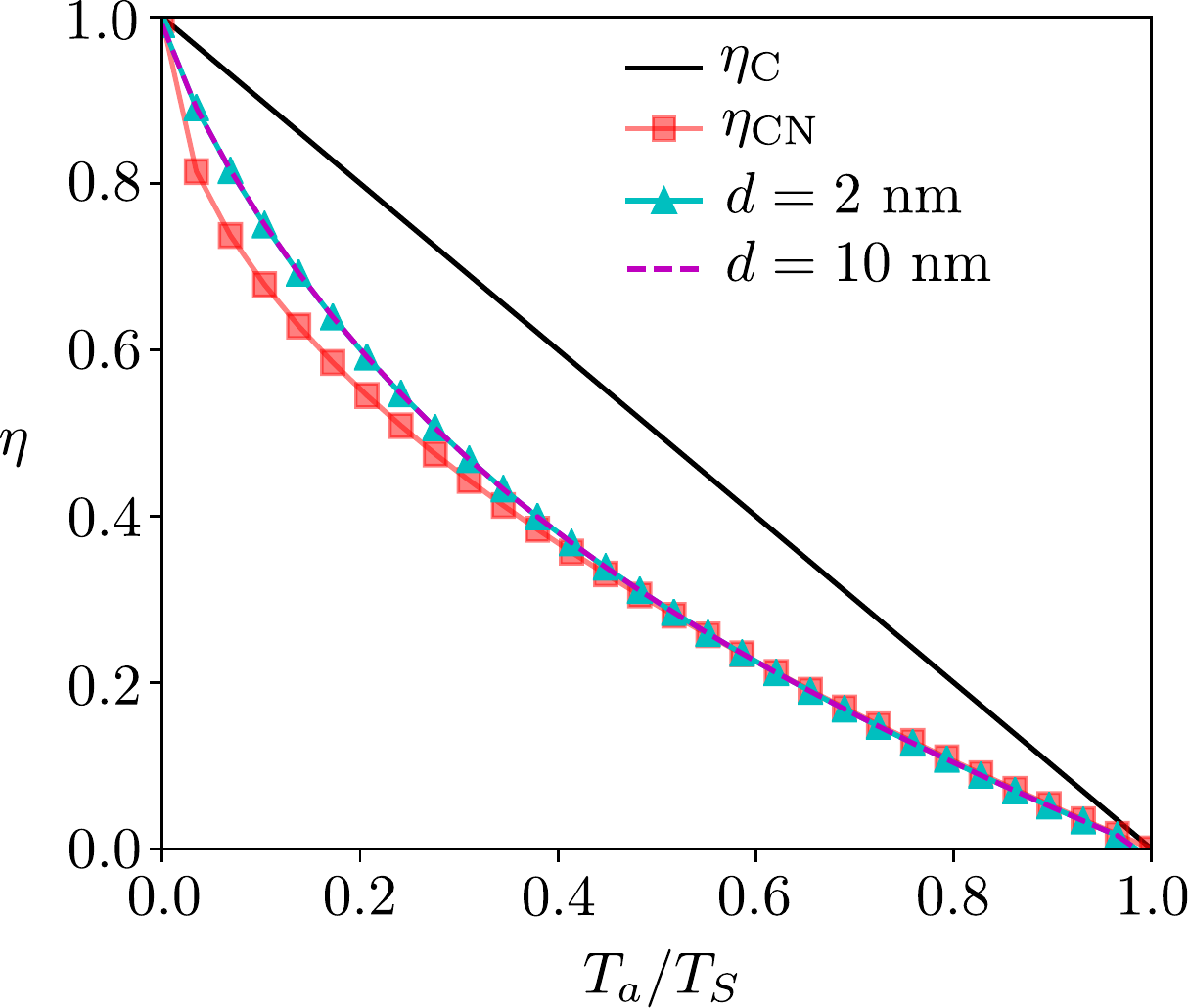}
	\caption{Efficiency $\eta$ as a function of $T_a/T_S$ considering $ \delta = 3 $ meV and barrier widths $d =2$ (blue triangles and solid line) and $d = 10$ nm (purple dashed line). The upper limit is the efficiency of Carnot, $\eta_{\mathrm{C}}$, (black solid line) while the lower limit is given by Chambadal-Novikov efficiency, $\eta_{\mathrm{CN}}$ (red squares and solid line).}
	\label{Fig5}
\end{figure}

In classical thermodynamics, heat engines are often analyzed without taking into account the irreversibleness of real engines. This results in the prediction of an unreachable limit for classical thermal machines, known as the Carnot efficiency, defined as $\eta_{\mathrm{C}}=1-T_a/T_S$. However, endoreversible thermodynamics makes more realistic assumptions about heat transfer and provides a new upper limit for thermal efficiencies: the operating efficiency of a semi-ideal classical thermal engine at maximum power~\cite{dorfman2018,chambadal1957,novikov1958,curzon1975}, also known as Chambadal-Novikov efficiency $ \eta_{\mathrm{CN}} = 1 - \sqrt{T_a/T_S} < \eta_{\mathrm{C}} $. In Fig. \ref{Fig5}, both limits are compared with the behavior of the EMP of the QDM, considering $ \delta = 3 $ meV, for barrier widths with values $d = 2$ nm (blue triangles and solid line) and $d = 10$ nm (purple dashed line). The results show the QDM exceeds the Chambadal-Novikov limit (red squares and solid line) for temperature ratio $T_a/T_S < 0.5$, and the behavior does not depend on the value of $d$. Still, the device stays below the Carnot efficiency (black solid line). Despite the similarity between the efficiency of the system of interest and the limit of Chambadal-Novikov, the physics is quite different if compared with other proposals~\cite{esposito2010, dorfman2018} found in literature, once the treatment of the present proposal analyzes the device using the steady-state solution of a quantum master equation, instead of considered as a four-stroke engine.

To summarize, it is used the master equation approach to study the interplay between quantum coherence and the manipulation of energy offsets over the behavior of the efficiency at maximum power of a quantum photocell consisting of InAs/GaAs quantum dots coupled by coherent tunneling. Our proposal shows that a careful setup of offsets energy and barrier width creates conditions for high photovoltaic conversion, providing 30\% more of maximum power delivered in relation to the SQD.  In addition, the system exhibits gain, if compared to the efficiency of Chambadal-Novikov at maximum power, although is still limited by the efficiency of Carnot, in agreement with the second law of thermodynamic.

\begin{acknowledgments}
This work was supported by CAPES, and the Brazilian National Institute of Science and Technology of Quantum Information (INCT-IQ), grant 465469/2014-0/CNPq.
\end{acknowledgments}

\end{document}